\documentclass[runningheads]{llncs}
\usepackage[T1]{fontenc}
\usepackage{graphicx}
\usepackage{amssymb}
\usepackage{amsmath}
\usepackage{subcaption}

\begin{document}
\title{Blockchain-based Federated Recommendation with Incentive Mechanism}
%
%
\newcommand{\equalcontrib}{\textsuperscript{\textasteriskcentered}}
\newcommand{\correspondingauthor}{\textsuperscript{\textdagger}}

\author{
Jianhai Chen\inst{1}\equalcontrib\and
Yanlin Wu\inst{1}\equalcontrib\and
Dazhong Rong\inst{1} \and
Guoyao Yu\inst{1} \and
Lingqi Jiang\inst{1} \and
Zhenguang Liu\inst{1}\and
Peng Zhou\inst{2}\and
Rui Shen\inst{1}\correspondingauthor}
\authorrunning{J. Chen, Y. Wu, et al.}
%
\institute{College of Computer Science and Technology, Zhejiang University, China\\
\email{\{chenjh919,3200101006,rdz98,yuguoyao23,3210100336,rshen\}@zju.edu.cn}\\
\email{liuzhenguang2008@gmail.com}
\and
Info \& Comm Branch, State Grid Zhejiang Electric Power Company, China\\
\email{zhou\_peng@zj.sgcc.com.cn}}
\maketitle              

\renewcommand\thefootnote{}
\footnotetext{\equalcontrib The two authors contributed equally to this work.}
\footnotetext{\correspondingauthor Corresponding author.}
\renewcommand\thefootnote{\arabic{footnote}}

\begin{abstract}
Nowadays, federated recommendation technology is rapidly evolving to help multiple organisations share data and train models while meeting user privacy, data security and government regulatory requirements. However, federated recommendation increases customer system costs such as power, computational and communication resources. Besides, federated recommendation systems are also susceptible to model attacks and data poisoning by participating malicious clients. Therefore, most customers are unwilling to participate in federated recommendation without any incentive. To address these problems, we propose a blockchain-based federated recommendation system with incentive mechanism to promote more trustworthy, secure, and efficient federated recommendation service. First, we construct a federated recommendation system based on NeuMF and FedAvg. Then we introduce a reverse auction mechanism to select optimal clients that can maximize the social surplus. Finally, we employ blockchain for on-chain evidence storage of models to ensure the safety of the federated recommendation system. The experimental results show that our proposed incentive mechanism can attract clients with superior training data to engage in the federal recommendation at a lower cost, which can increase the economic benefit of federal recommendation by 54.9\% while improve the recommendation performance. Thus our work provides theoretical and technological support for the construction of a harmonious and healthy ecological environment for the application of federal recommendation.

\end{abstract}
\section{Introduction}
Nowadays, various application platform service providers (\textit{e.g.,} State Grid APP) have formed a ``data silo'' after a long period of accumulation. Thus if there is a mechanism that can open up the silo to make the data flow, it will bring a great deal of value. Utilizing data for recommendations is a primary way of maximizing its potential. However, the volume and quality of data play a decisive role in the effectiveness of recommendation algorithms~\cite{DBLP:conf/ksem/HuRCHL23}. At present, most commercial enterprises face the problems of insufficient data quantity and poor data quality~\cite{nasr2019comprehensive,qu2020decentralized,DBLP:journals/corr/abs-2403-15010}, which makes it difficult to fully develop high-quality recommender system technology. Furthermore, global attention to data privacy protection and security has led to the issuance of legislations like the most notably General Data Protection Regulation (GDPR)\cite{yang2019federated,regulation2016regulation}, which have made data acquisition even more challenging. In this context, federated recommendation was born. It is a federated learning framework that enables different organizations share data and train recommendation models together, balancing the demands of data security, user privacy concerns, and government regulations. A typical federated recommendation system is composed of a central server and other clients that provide training data. The server releases federated recommendation assignments and manages clients' participation in the process of model training. Clients upload local model parameters or gradients, which the server aggregates. Then the server sends global recommendation model parameters back to all participating clients.

However, federated recommendations tend to increase client system costs, including power, communication resources and computing resources. Besides, customers are exposed to privacy leakage risks during the training process~\cite{ma2020safeguarding}. Federated recommendation systems are also susceptible to model attacks and data poisoning by the participants. Thus, most clients are unwilling to engage in federated recommendations voluntarily without any economic incentive. To address these drawbacks, we propose a blockchain-based federated recommendation system with a truthful incentive mechanism that can be divided into three phases: 1) Clients get the opportunity to participate in training through auction. 2) Selected clients train and upload local model parameters for evidence storage on the blockchain. 3) The server uses uploaded model parameters on the chain to aggregate. This mechanism enables the server to retrace user-uploaded model information at any time, effectively tracking down malicious behaviours while credibly calculating incentives to be paid out to honest users.  

The main contributions of this paper can be summarized as follows. 

\begin{enumerate}
    \item We construct a federated recommendation system based on NeuMF~\cite{he2017neural} and FedAvg~\cite{mcmahan2017communication}, which keeps the user's embedding layers locally to protect user privacy data.
    \item We propose a D3QN-based reverse auction mechanism by which clients need to bid for the opportunity to participate in federated recommendations. Through our auction mechanism, we can select the optimal client set to achieve the highest economic benefits and save the costs.
    \item We employ blockchain for on-chain evidence storage of models to ensure the safety and the trustworthiness of the federated recommendation system.
\end{enumerate}
\section{Related work}
The research on federated recommendation systems is still in its early stages compared to conventional centralized recommendation systems.
Recently, Lin et al.~\cite{lin2023federated} used various recommendation algorithms as microservices and proposed FedNeuMF. FedNeuMF is a federated collaborative filtering recommendation model, accomplishing distributed recommendation microservices application which preserves privacy. However, these federated recommendation systems are neither truthful and secure nor do they provide incentives for the participants. 

Stealing or modifying client-uploaded local model parameters is a major problem in federated recommendations~\cite{rong2022fedrecattack,DBLP:conf/ijcai/RongHC22,DBLP:conf/icde/Zhang0RZ0S24}. Besides, if the server is attacked, it can potentially incur a single-point failure and thereby impair the entire federated recommendation system. As a solution, Blockchain, a ledger technology~\cite{nguyen2021bedgehealth}, can be adapted to offer trust assurance for federated recommendations due to its unique characteristics of immutability, decentralization and traceability.

Although there has been some research on blockchain-based federated learning~\cite{feng2021bafl,qin2021privacy,issa2023blockchain,patel2020kirti,himeur2022blockchain,liu2023efficient}, there is still very little research on blockchain-based federated recommendation. Besides, blockchain-based federation learning is not applicable to the actual scenario of federation recommendation. In traditional blockchain-based federation learning, the client uploads complete model parameters to the server for aggregation, and the server uploads complete global model parameters to the client. However, in the federated recommendation scenario, due to factors such as protecting the client's user privacy and network efficiency, clients only upload model parameters excluding user embedding layers instead of complete parameters. Therefore, we propose a new truthful and secure blockchain-based federated recommendation framework with incentive mechanism~\cite{chen2019truthful,chen2020truthful}.
\section{Preliminaries}
We first introduce the preliminary knowledge of federated recommendation, then present the model on which our federated recommendation system is based.

\subsection{Federated Recommendation}
Current federated recommendation systems typically base on a Client-Server architecture, where a group of edge devices known as clients take part in training under the direction of a central server. Each client $k$ has its own dataset $D_k$ , containing $M$ users and $N$ items. The dataset can be represented as an user-item interaction matrix $\mathbf{Y}\in\mathbb{R}^{M\times N}$. In explicit feedback, value $y_{ui}$ within this matrix represents the rating of user $u$ for item $i$. In implicit feedback, $y_{ui}$ equals 1 if the user $u$ once interacted with the item $i$, otherwise $y_{ui}$ equals 0. Each client's local learning can be formalized as learning $\hat{y}_{ui}=f(u,i|\mathbf{w}_k)$, where $f$ refers to the function that maps model input to the predicted score, $\hat{y}_{ui}$ is the predicted rating of user $u$ for item $i$ and $\mathbf{w}_k$ indicates the model parameters of client. After the local training converges, each client uploads its local model parameters to the server for model aggregation. The server only gathers model parameters rather than local training data from the clients, reducing data transmission cost and at the same time securing users' privacy. The process of Federated Recommendation contains the following three steps:

\textbf{Step 1: Initialization.} The server initializes the global model parameters $\mathbf{w}_g$ and transmits them to all clients, to serve as initial model parameters.

\textbf{Step 2: Local model training and updates.} After receiving the global parameters $\mathbf{w}_{g}^{t}$, where $t$ signifies the latest global update epoch, local client $k$ proceeds training through local dataset $D_k$ until it converges, i.e., 
\begin{equation} 
\mathbf{w}_k^{t^*}=\arg\min_{\mathbf{w}_k^t}L(\mathbf{w}_k^t),
\end{equation}
where $L$ represents the local model's loss function. Then, each client uploads the updated local model parameters $\mathbf{w}_{k}^{t}$ to the server.

\textbf{Step 3: Global model aggregation.} The server aggregates all received local model and sends the updated global model parameters $\mathbf{w}_{g}^{t+1}$ to all clients.

Steps 2 and 3 are iterated until the global model converges, \textit{i.e.,} the global loss function $L(\mathbf{w}_g^t)$ is minimised. $L(\mathbf{w}_g^t)$ can be formulized as: 
\begin{equation} 
L(\mathbf{w}_g^t)=\frac{1}{\mathrm{N}}\sum_{k\in\mathcal{N}}L(\mathbf{w}_k^t),
\end{equation}
where $\mathcal{N}$ denotes the set of local clients and $\mathrm{N}$ denotes set size.

\subsection{Based Recommendation Model}
We use the NeuMF model proposed in \cite{he2017neural} and federate it. NeuMF consists of two parts: one is the Generalized Matrix Factorization (GMF) which learns the linear patterns of user-item interactions, the other is a Multi-Layer Perceptron (MLP) which learns the non-linear patterns of user-item interactions. The outputs of these two parts are then merged into a predicted score. The NeuMF model can be expressed as the following formulation: 
\begin{equation}  
\begin{aligned} \phi^{GMF}& =\mathbf{p}_u^G\odot\mathbf{q}_i^G, \\ \phi^{MLP}& =a_L(\mathbf{W}_L^T(a_{L-1}(...a_2(\mathbf{W}_2^T\begin{bmatrix}\mathbf{p}_u^M\\\mathbf {q}_i^M\end{bmatrix}+\mathbf{b}_2)...))+\mathbf{b}_L), \\ \hat{y}_{ui}& =\sigma(\mathbf{h}^{T}\begin{bmatrix}\phi^{GMF}\\\phi^{MLP}\end{bmatrix}), \end{aligned}
\end{equation}
where $\phi$ denotes the output, $\odot$ indicates the element-wise product, $\mathbf{p}_u^G$ and $\mathbf{p}_u^M$ represent the user latent vectors in GMF and MLP, and $\mathbf{q}_i^G$ and $\mathbf{q}_i^M$ similarly represent the item latent vectors in GMF and MLP, which are the results of multiplying the user and item embedding layers $\mathbf{P}$ and $\mathbf{Q}$ by the one-hot coding of the user and item. $\mathbf{W}_n$, $a_{n}$ and $\mathbf{b}_n$ denote the weight matrix, activation function and bias vector of the $n$-th layer perceptron respectively. $\sigma$ represents the sigmoid function and $\mathbf{h}$ denotes the edge weight of the output layer.

\section{Our Method}
In this chapter, we first propose a reverse auction mechanism designed for Federated Recommendation. Through this mechanism, we can select the optimal set of clients to maximize social surplus, enhancing recommendation performance while saving cost as we have fewer clients to pay for. We then introduce the process of training and updating the local recommendation model. Lastly, we present a mechanism for aggregating global models.

\subsection{The Reverse Auction Scheme}
In Federated Recommendation, the parameters of the global recommendation model are obtained by aggregating all uploaded local recommendation model parameters. Therefore, the effectiveness of local model training is crucial to the overall performance of the recommendation system. In practical recommendation scenarios, data possessed by different clients is often imbalanced and non-iid, and the heterogeneity of data substantially slows down the convergence speed of Federated Recommendation and undermines its final recommendation effect. Thus, it is necessary to make a preliminary judgment on data quality from clients before they participate in training to exclude those with extremely low data quality. We employ Earth Mover's Distance (EMD) to signify the data quality of clients. For client $i$, the EMD distance $\sigma_i$ can represent the difference in data distribution between client $i$ and global, i.e.,
\begin{equation} 
\sigma_i=\sum_{r\in\mathcal{R}}\|\mathbb{P}_i(y=r)-\mathbb{P}_g(y=r)\| ,
\end{equation}
where, $\mathcal{R}$ denotes the range of ratings, $\mathbb{P}_i$ denotes the local data distribution, $\mathbb{P}_g$ denotes the global data distribution obtained empirically.

\begin{figure}[ht]
    \centering
    \includegraphics[width=0.9\linewidth]{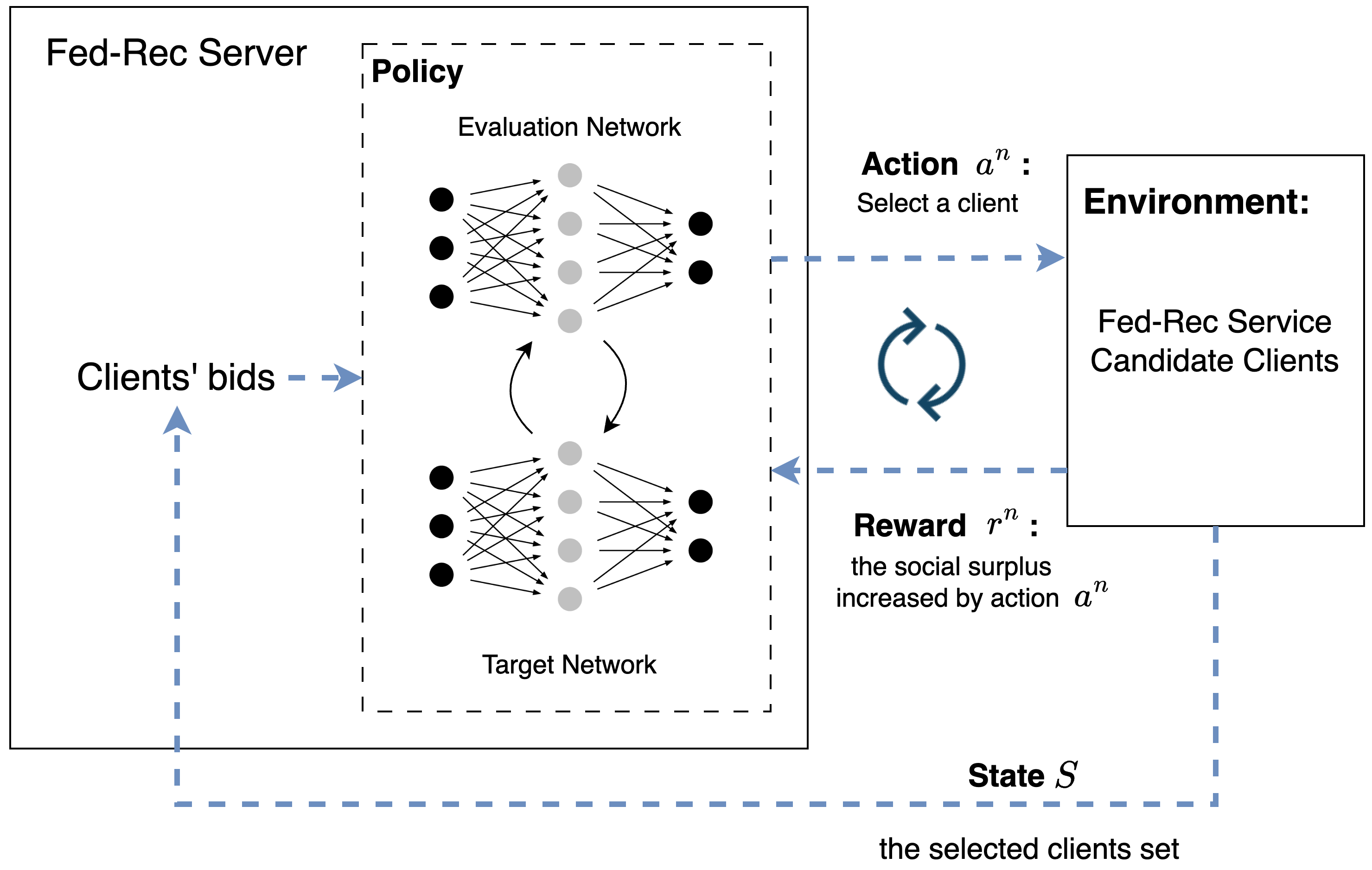}
    \caption{The best clients select mechanism based on D3QN.}
    \label{fig:D3QN}
\end{figure}

The goal of our designed reverse auction mechanism is to maximize the social surplus of federated recommendation, where the profit generated by the federated recommendation is determined by the final global model's accuracy. Before training, we can estimate the data utility of client $i$ based on the size of the client's training data $s_i$ and the EMD distance $\sigma_i$ of the training data set. \cite{jiao2020toward} proposes that the relationship between the final global model quality $Q$ of federated learning and the total dataset size $D$ of the participating training clients' set $C$ and the average EMD distance $\Delta$ can be represented as follows: 
\begin{equation} 
Q(C)=\alpha(\Delta)-\kappa_{1}\mathrm{e}^{-\kappa_{2}(\kappa_{3}D)^{\alpha(\Delta)}},
\end{equation}
where $D$ is the total dataset size of the participating training clients, i.e., $D(C) = \sum_{i\in C}s_i$. $\Delta$ is the average EMD distance of the participating training clients, i.e., $\Delta(C) = \frac{\sum_{i\in C}\sigma_{i}}{|C|}$. We specify $\Delta(\emptyset) = 0$. $\alpha(\Delta) = \kappa_{4}\exp(-(\frac{\Delta+\kappa_{5}}{\kappa_{6}})^{2}) < 1$. The machine learning quality function $Q(C)$ is determined by a curve fitting method. $\kappa_1,\ldots,\kappa_6 > 0 $ are positive curve fitting parameters. Since the revenue $R$ of the system is related to the quality of global model, $R$ can be represented as $R(C)=\lambda Q(C)$, where $\lambda$ denotes the satisfaction weight parameter. Ignoring power expenditure and communication costs, the cost of our federated recommendation system considers only the incentives paid to the clients. In our reverse auction mechanism, client $i$ needs to send its local dataset size $s_i$, EMD distance $\sigma_i$ and bid $b_i$ to the server. Thus, the system social surplus can be calculated as $S=R(C)-\sum_{i=1}^{n}b_{i}C_{i}$, where we express the client selection set $C$ in vector form, i.e., $C\in(0,1)^n$, where $C_i=1$ means client $i$ is selected, and $C_i=0$ means client $i$ is not selected. Therefore, the goal of our reverse auction mechanism is to find the best set of clients to maximize our system's social surplus. The optimal client selection vector can be expressed as follows: 
\begin{equation} 
C^{*}=\mathrm{argmax}_{{C\in(0,1)^{n}}}\left[R(C)-\sum_{i=1}^{n}b_{i}C_{i}\right].
\end{equation}
Due to the NP-hard nature of maximizing social surplus, we use the Deep Reinforcement Learning(DRL) algorithm Double Dueling Deep Q-Network (D3QN) to calculate $C^{*}$. This model combines Double Q-Learning with Dueling Network architecture into Deep Q-Networks, improving learning performance and stability, enabling it to avoid overestimation and handle more complicated tasks. Our proposed D3QN framework is shown in Fig.~\ref{fig:D3QN}, comprising states, actions, state transitions, rewards, and policy defined as follows.

\textbf{State:} A state at step $n$ can be formalized as $s^{n}=\{s_{1}^{n},...,s_{i}^{n},...,s_{N}^{n}\}$ ,which contains the states of clients representing whether they have been selected into the candidate set $\mathcal{V}^{n}$. That is, if $i\in\mathcal{V}^{n}$ then $s_{i}^{n}=1$, otherwise $s_{i}^{n}=0$. 

\textbf{Action:} An action $a^n$ is an index, indicating that at step $n$, client $a^n$ is selected into the candidate list $\mathcal{V}^{n+1}$ from not being a candidate item in $\mathcal{V}^{n}$. 

\textbf{State transition:} If action $a^n$ is taken, the global state would transit from $s^{n}$ to $s^{n+1}$ as the client's state $s_{a^n}$ would transit from 0 to 1. Besides, client $a^n$ would be added to the candidate set, i.e., $\mathcal{V}^{n+1}=\mathcal{V}^{n}\cup\{a^{n}\}$. 

\textbf{Reward:} The reward $r^n$ is the increase of social surplus when action $a^n$ is taken at state $s^n$. It can be calculated by the following formula: 
\begin{equation}  
r^n=S(\mathcal{V}^n\cup\{a^n\})-S(\mathcal{V}^n),
\end{equation}  
where $S$ is the social surplus function defined by equation (6). 

\textbf{Policy:} We first use evaluate network $Q^{\prime}(s^{n},a^{n};w_{e})$ to select the optimal action. Then we use the target network $Q(s^{n},a^{n};w_{t})$ to compute the value of this action's to obtain target value. The combination of these two networks can avoid overestimation. Besides, we choose the $\epsilon-greedy$ strategy, expressed as follows:  
\begin{equation}  
a^n=\begin{cases}
\underset{a^n}{\operatorname*{\mathrm{argmax}}}Q(s^n,a^n;w_t)& \text{, with a probability } \epsilon \\
\text{randomly select a client in } \mathcal{N} \setminus \mathcal{V}^n & \text{, with a probability } 1-\epsilon
\end{cases}.
\end{equation} 

\subsection{The Updating Process of Local Model}
We have introduced our based recommendation model, the NeuMF model, in Section 3.2. We can summarise the model parameters of NeuMF as user embeddings $\mathbf{p}$, item embeddings $\mathbf{q}$ and the parameters $\mathbf{W}$, $\mathbf{b}$, $a$ of the perceptual layers. In classical federated learning, the client directly downloads the complete global model parameters as local model initialisation in each round, and uploads the complete local model parameters after training is finished. However, in our federated recommendation scenario, in order to protect the client user's private information and improve the network efficiency, the client generally does not upload the user's embeddings $\mathbf{p}$. Thus, only $\mathbf{q}$, $\mathbf{W}$, $\mathbf{b}$, $a$ are uploaded to the server for aggregation. The local model updating process of clients in our federated recommendation system is as follows.

\textbf{Step 1:} Local model initialization. Download the latest global recommendation model parameters $\mathbf{w}_g$ from the blockchain. The parameter $\mathbf{w}_g$ here contains only global item embeddings $\mathbf{q_g}$ and the parameters $\mathbf{W_g}$, $\mathbf{b_g}$, $a_g$ of the global perceptual layers, without user embeddings $\mathbf{p}$. Local model parameter $\mathbf{w}_k$ contains local item embeddings $\mathbf{q_k}$ and the parameters $\mathbf{W_k}$, $\mathbf{b_k}$, $a_k$ of the local perceptual layers as well as the local user embeddings $\mathbf{p_k}$. Therefore, $\mathbf{q_k}$, $\mathbf{W_k}$, $\mathbf{b_k}$, $a_k$ are overwritten by $\mathbf{w}_g$, while the user embeddings $\mathbf{p_k}$ remain unchanged.

\textbf{Step 2:} Local model training. The model conducts a forward propagation first, obtaining the predicted score $y_{ui}$. The loss is then computed and the objective function used can be expressed as: 
\begin{equation}  
L=\sum_{(u,i)\in\mathcal{Y}\cup\mathcal{Y}^-}w_{ui}(y_{ui}-\hat{y}_{ui})^2, 
\end{equation}
where $\mathcal{Y}$ signifies all observed interactions in the user-item interaction matrix $\mathbf{Y}$, and $\mathcal{Y}^-$ denotes all negative samples, which are sampled from interactions that have not been observed. $w_{ui}$ represents the weight of the training instance $(u,i)$. The model parameters can then be updated through standard back-propagation. The parameter update expression for the $t$-th epoch is as $\mathbf{w}_k^{t}=\mathbf{w}_k^{t-1}-\alpha\nabla\!L$, where $\alpha$ represents the learning rate. Finally, the model iterates and optimizes until the loss no longer decreases significantly or the pre-set maximum number of iterations is reached.

\textbf{Step 3:} Local model upload. After the client completes local training, the final local model parameters $\mathbf{w}_k^*$ are uploaded to the record blockchain. The blockchain enables server to retrace user-uploaded model information at any time, effectively tracking down malicious behaviours
while credibly calculating incentives through records.

\subsection{The Design of Aggregation Mechanism}
We perform an aggregation operation with the FedAvg algorithm, which is represented as $\mathbf{w}_{g}^{t+1}=\sum_{k=1}^N\frac{n_k}n\mathbf{w}_{k}^t$, where $N$ represents the number of honest clients, $n$ represents the size of the sum of all datasets belonging to clients, $n_k$ refers to the size of the dataset of the client $k$, $\mathbf{w}_{k}^t$ represents the model parameters for the client $k$ in global epoch $t$, and $\mathbf{w}_{g}^{t+1}$ stands for the global model parameters for $(t+1)$-th global epoch. After the completion of aggregation, the latest global model parameters are uploaded to the blockchain while the selected clients are paid based on the previous bids. The blockchain records the global model information, ensuring that the records of the auction process and the training process are tamper-proof and traceable. This ensures the security and trustworthiness of the federated recommender system.
\section{Experiments}
In this section, we display our experimental setting, experimental methodology, and performance evaluation. We first introduce the simulation setup for various clients, then test the auction benefits of our reverse auction mechanism.
\begin{table}[b]
    \centering
    \caption{Experimental environment}
    \begin{tabular}{|l|l|}
    \hline
    \textbf{Component} & \textbf{Specification} \\ \hline
    CPU & 18 vCPU AMD EPYC 9754 128-Core Processor \\ \hline
    GPU & RTX 4090D(24GB) * 1 \\ \hline
    OS & Ubuntu 20.04 \\ \hline
    Cuda & 11.3 \\ \hline
    Pytorch & 1.11.0 \\ \hline
    Python & 3.8 \\ \hline
    \end{tabular}
    \label{tab:my_label}
\end{table}

\subsection{Experimental Settings}
\textbf{Experimental environment.} Table~\ref{tab:my_label} shows our experimental environment.

\textbf{Dataset.}
We experimented with MovieLens dataset. The MovieLens dataset is a publicly available film rating dataset released by the GroupLens laboratory. It's widely applied in research related to recommendation systems. The dataset includes versions with various amounts of ratings and we used the MovieLens 1M version. The users' ratings range from 1 to 5, denoting from strongly disliked (1) to strongly liked (5). The dataset also offers timestamp information for each rating, allowing researchers to investigate rating patterns based on time.

\textbf{Evaluate Protocols.}
To authentically simulate various types of clients in the setting of federated recommendation scenarios, we set up a total of 20 clients with different volume and quality of train data. To analyze the performance of our proposed global model aggregation mechanism and reverse auction mechanism, we performed experiments through leave-one-out cross-validation. For each user, we retained their latest interaction as the test set and used the other data as the training set. Since sorting all the user ratings is time-consuming we adopted a common used strategy where we randomly selected 50 items for each user that had not been interacted with and then ranked the items in our test set among them. We measure the global model's rating prediction performance through the Mean Squared Error (MSE) loss between the predicted rating of the global model and the test set's real rating. Additionally, we also employ Hit Ratio (HR) and Normalized Discounted Cumulative Gain (NDCG) to analyze the performance of the ranked list. We only consider the top 10 of the ranked list. HR shows whether the test items appear within the top 10, while NDCG accounts for the hit location by giving higher scores to more highly ranked hits.

\textbf{Parameter Settings.}
In our experiments, we implemented the federated recommendation model by Pytorch, using FedAvg for the global model aggregation and the Adam algorithm for parameter optimization. During the training, each training instance consists of one positive instance and four negative instances. We adopted an MSE loss function to train our model. For NeuMF, we initialized the parameters of the user and item embedding layers with a Gaussian distribution (mean value 0, variance of 0.01). We used Xavier Uniform to initialize the parameters of the multi-layer perceptron and Kaiming Uniform to initialize the parameters of the output layer. We set $\kappa1 = 0.361$, $ \kappa2 = 4.348$, $ \kappa3 = 10^{-3}$, $ \kappa4 = 0.993$, $ \kappa5 = 0.31$ and $\kappa6 = 1.743$ due to \cite{jiao2020toward}. Besides, we set learning rate to $0.0005$, training batch size to $64$, the size of predictive factors to $32$, the number of local iteration epochs to $5$, the number of global iteration epochs to $10$ and the satisfaction weight parameter $\lambda = 3000$.

\begin{table}[t]
    \centering
    \caption{Comparison of HR@10 and NDCG@10 under different mechanisms.}
    \begin{tabular}{|c|c|c|}
    \hline
    \textbf{Method} & \textbf{HR@10} & \textbf{NDCG@10} \\ 
    \hline
    Greedy-All & 0.511 & 0.290 \\ 
    \hline
    Simple-Auction & 0.525 & 0.288\\ 
    \hline
    D3QN-Auction & 0.543 & 0.317 \\ 
    \hline
    \end{tabular}
    \label{tab:comparison}
\end{table}
\begin{figure}[t]
    \begin{subfigure}{.5\textwidth}
    \centering
    \includegraphics[width=\linewidth]{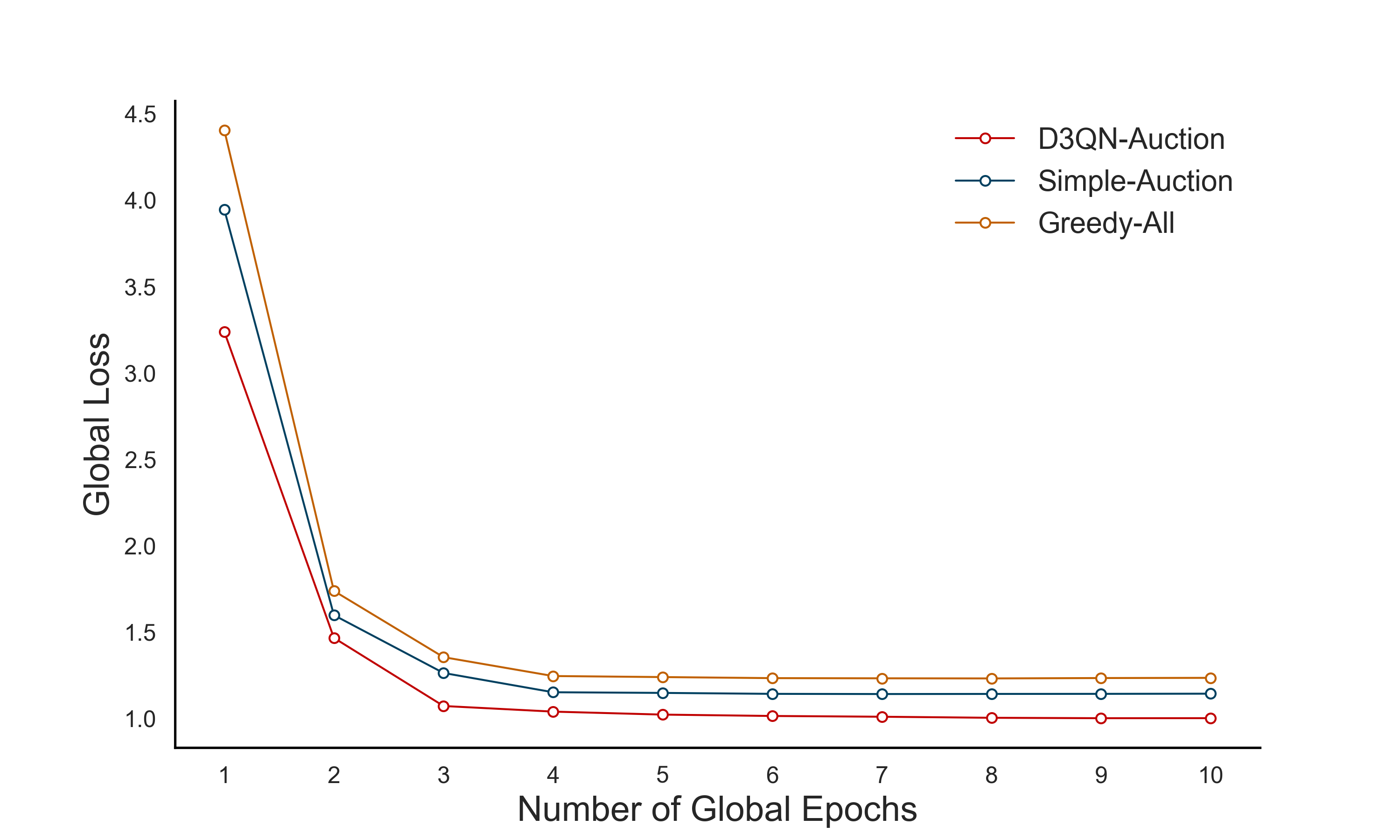}
    \caption{global loss}
    \label{fig:poison}
    \end{subfigure}
    \begin{subfigure}{.45\textwidth}
    \quad
    \centering
    \includegraphics[width=\linewidth]{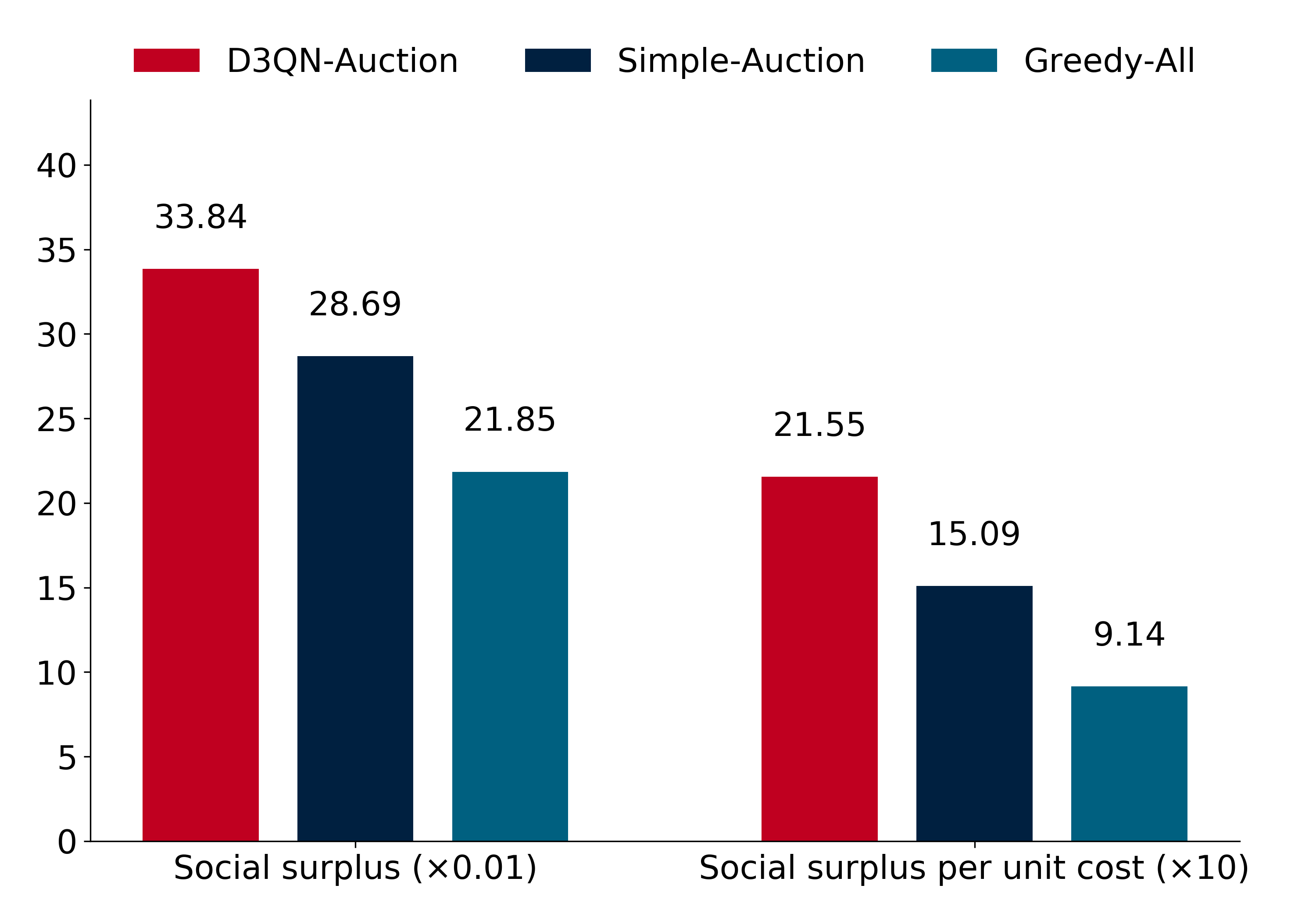}
    \caption{total social surplus and per-unit cost social surplus}
    \label{fig:social_remain}
    \end{subfigure}
    \caption{Comparison under different auction mechanisms.}
\end{figure}

\subsection{Performance of Reverse Auction}
In this section, we analyse the performance of the reverse auction mechanism based on D3QN. We employ our proposed reverse auction mechanism to select the optimal client set that maximizes social surplus and compare our reverse auction mechanism with two other auction mechanisms. One is Simple-Auction, which prioritizes clients with lower bid prices and selects 80\% of all clients. The other is Greedy-All, which selects all clients. Our D3QN-Auction selected an optimal client set consisting of 11 clients, the Simple-Auction mechanism and the Greedy auction mechanism chose 16 and 20 clients, respectively. Fig.~\ref{fig:poison} shows the comparison of global loss under three auction mechanisms, reflecting the global model's prediction accuracy. It is evident that the global loss of our proposed D3QN-Auction algorithm is lowest, reaching only 1.0023. This suggests that our reverse auction mechanism can achieve better training outcomes with fewer clients. Besides, the D3QN-Auction allows the global model to converge faster, thus it can save more time and resources.
Table~\ref{tab:comparison} shows the comparison of the HR@10 and NDCG@10 metrics of the global recommendation model under different auction mechanisms. It illustrates that, in terms of recommendation hit rates and ranking accuracy, D3QN-Auction still surpasses Greedy-All and Simple-Auction.
Fig.~\ref{fig:social_remain} presents a comparison of the total social surplus and per-unit cost social surplus under different auction mechanisms. In terms of total social surplus, our proposed D3QN-Auction method reached the highest value of 3384, which exceeds Greedy-All by 54.9\%. In terms of per-unit cost social surplus, D3QN-Auction also achieved the best outcome, reaching  2.155, outperforming Greedy-All by 135.8\%. Therefore, compared to Simple-Auction and Greedy-All, our proposed D3QN-based reverse auction mechanism can provide greater economic benefits and a higher return-on-investment ratio as it can achieve the expected recommendation effect at a lower cost. Under such an incentive mechanism, more high-quality clients can be attracted to join our federated recommendation tasks, leading to better recommendation results.

\section{Conclusion and Future Work}
In this paper, we primarily focuses on the research of blockchain-based federated recommendation with incentive mechanism.
Our proposed blockchain-based federated recommendation system solves the problem of verifiable client contribution and the trustworthiness of the incentive mechanism. 
In the future, we consider adding some client malicious behavior defence mechanisms.

\begin{credits}
\subsubsection{\ackname} This work was funded by the Key R\&D Program of Zhejiang Province (No. 2022C01086 and No. 2023C01217) and the National Natural Science Foundation of China (No. 62372402).
\end{credits}
%
%
%
\bibliographystyle{splncs04}
\bibliography{main}
\end{document}